\documentclass[prl,twocolumn,aps,showpacs,superscriptaddress]{revtex4}
\usepackage{graphics}
\usepackage{epsfig}
\usepackage{amsmath}
\usepackage{amssymb}
\usepackage{color}
\usepackage{hyperref}

\newcommand{\rom}[1]{\uppercase\expandafter{\romannumeral #1\relax}}

\begin{document}
\newcommand{\beq}{\begin{equation}}
\newcommand{\eeq}{\end{equation}}

\title{Origami Multistability:  From Single Vertices to Metasheets}

\author{Scott Waitukaitis}
\affiliation{Huygens-Kamerlingh Onnes Lab, Leiden University, PObox 9504, 2300 RA Leiden, The Netherlands}
\author{R\'emi Menaut}
\affiliation{Huygens-Kamerlingh Onnes Lab, Leiden University, PObox 9504, 2300 RA Leiden, The Netherlands}
\affiliation{\'Ecole Normale Sup\'erieure de Lyon, Université Claude Bernard Lyon 1, BP 7000 69342 Lyon Cedex 07, France}
\author{Bryan Gin-ge Chen}
\affiliation{Instituut-Lorentz for Theoretical Physics, Leiden University, PObox 9506, 2333 CA Leiden, The Netherlands}
\author{Martin van Hecke}
\affiliation{Huygens-Kamerlingh Onnes Lab, Leiden University, PObox 9504, 2300 RA Leiden, The Netherlands}

\date{\today}

\begin{abstract} We explore the surprisingly rich energy landscape of origami-like folding planar structures. We show that the configuration space of rigid-paneled degree-4 vertices, the simplest building blocks of such systems, 
consists of at least two distinct branches meeting at the flat state. This suggests that generic vertices are at least bistable, but we find that the  nonlinear nature of these branches allows for vertices with as many as five distinct stable states. In vertices with collinear folds and/or symmetry, more branches emerge leading to up to six stable states.  Finally, we introduce a procedure to tile arbitrary 4-vertices while preserving their stable states, thus allowing the design and creation of multistable origami metasheets.

\end{abstract}

\maketitle

Mechanical metamaterials are elastic media with extraordinary properties that arise from their microstructure \cite{Kadic:2008,Lakes_science1987, Milton:1992, Kadic:2012,Buckman:2014, Lakes:2001,Nicolaou:2012,Mullin:2007,Bertoldi:2008,Bertoldi:2010,Overvelde:2012,Shim:2013,Florijn:2014,Kane:2014, Chen:2014}. Currently established functionalities include negative Poisson's ratio \cite{Lakes_science1987}, vanishing shear modulus \cite{Milton:1992, Kadic:2012,Buckman:2014},  negative compressibility \cite{Lakes:2001,Nicolaou:2012}, pattern transformation \cite{Mullin:2007,Bertoldi:2008,Bertoldi:2010,Overvelde:2012,Shim:2013}, switchable multistablility \cite{Florijn:2014} and topological insulation \cite{Kane:2014, Chen:2014}. While the building blocks for these materials are quasi-1D rods or springs, origami-inspired metamaterials made from folding planar structures open up new possibilities \cite{Miura:1985tt, Wei:2013kn, Schenk:2013kk, Silverberg:2014, Cheng:2014}.

Most recent attention has been focused on the Miura-ori, a fold tessellation well-known for its negative Poisson's ratio.  Silverberg {\it et al.}~recently used Miura-ori to create a metamaterial with tunable stiffness by introducing a reversible ``pop-through'' defect \cite{Silverberg:2014}.  This local defect, permitted via plate-bending, is one of a few specific examples of bistability in folding planar structures---others include the symmetric waterbomb vertex \cite{Hanna:2014} and the hypar \cite{Demaine:2011}.  Such multistability is a desirable property for the design of metamaterials as it allows reprogrammable reconfiguration of shape and bulk properties.  

\begin{figure}[t!]
\includegraphics[scale=0.95]{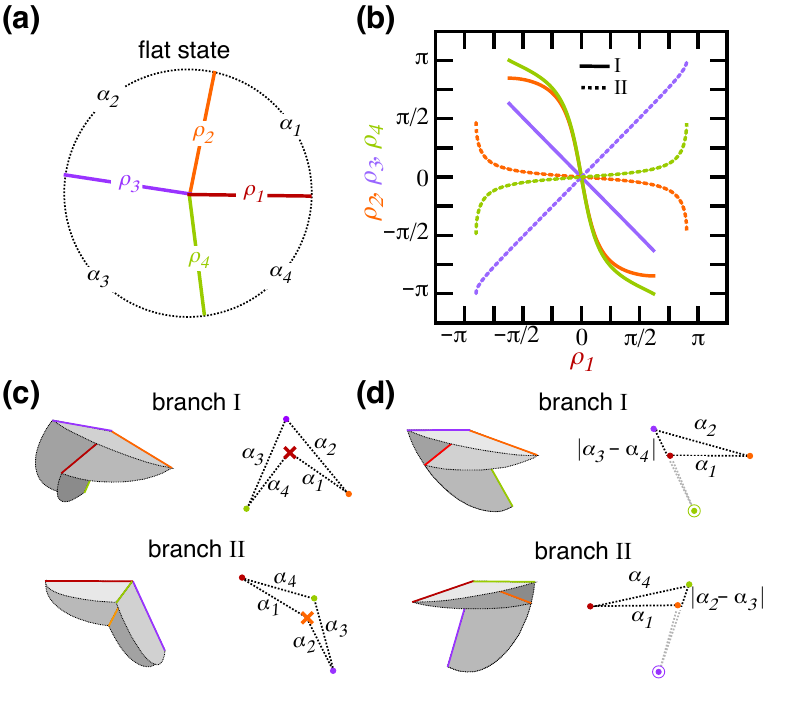}
\caption{Configuration space.  (a)  The flat state is defined by sector angles $\{\alpha_i\}$ (here $\{\alpha_i\}=\{ 1.4, 1.6, 1.9, 2\pi-4.9\}$) and a folded state by folding angles $\{\rho_i\}$. (b)   Configuration curves $\rho_{2,3,4}$ vs.~$\rho_1$ for branches I  and II.  (c)  3D renderings and schematic sideviews for folding motions on branch I and II (crosses designate unique folds). (d)  3D renderings and schematic sideviews for bindings on each branch (circled dots designate binding folds).   For movies illustrating the folding motions of each branch, see \cite{supplemental_material}.}
 \label{fig:vertex_geometry}
\end{figure}

Here we explore folding planar structures as a platform for globally reconfigurable metamaterials.  Degree-4 vertices are the simplest building blocks of such systems because they are one degree of freedom mechanisms \cite{cutnote1}.  We show that they are generically multi-branched mechanisms with {\it two} branches of folding motion emerging from the flat state [Fig.~\ref{fig:vertex_geometry}(b)].  We then study the energy landscape resulting from dressing each fold with a harmonic torsional spring. Generically, one expects such systems to be bistable (one minimum per branch), but we show that the nonlinear relations between folding angles [Fig.~\ref{fig:vertex_geometry}(b)] lead to complex energy landscapes with as many as five minima. We also demonstrate that tuning the fold energy parameters allows one to create monostable vertices, and we reveal why Miura-ori is typically just monostable. We then show how non-generic fold geometries can lead to more complex branch structures with up to six stable states. Finally, we illustrate a simple procedure to take any 4-vertex and encode all of its stability features into homogeneous stable states of a fold tessellation, thus creating multistable origami metasheets.

{\em Generic Configuration Space:}  We first explore the configuration space of generic 4-vertices, {\it i.e.}~those without collinear folds, symmetry or flat-foldability \cite{Demaine:2007}.  We specify the flat-state geometry by the set of sector angles $\{\alpha_i\}$, where each $\alpha_i < \pi$ and $\Sigma_i \alpha_i = 2\pi$  [Fig.~1(a)].  A folded state is described by the folding angles $\{\rho_i\}$, the complements of the dihedral angles between plates $i$ and $i-1$ (cyclic permutations understood). We take positive folding angles as ``valleys'' and negative ones as ``mountains.''  We begin by considering three basic questions:  {\em(i)} What are the possible mountain valley assignments for the folds? {\em(ii)} Which folds can be maximally folded to $\pm \pi$? {\em(iii)} What are the relationships between the folding angles? Below we summarize the answers; a detailed treatment, including non-generic folds and non-flat paper, is in preparation \cite{inprep}.

{\em(i)}  Huffman noted that one folding angle must have the opposite sign from the rest  \cite{Huffman:1976}.  Which folds can do this?  Fig.~\ref{fig:vertex_geometry}(c) shows two mountain-valley assignments for the vertex shown in panel (a).  In each case the ``unique'' fold with the sign opposite from the rest is "cupped" inside the others. This implies that fold $j$ can be unique if
\begin{equation}
\alpha_{j-1}+\alpha_j \le \pi,
\label{eq:odd_fold}
\end{equation}
where equality corresponds to a non-generic case \cite{inprep}.  For generic vertices it immediately follows that two folds are unique and straddle a common plate. This implies that generic 4-vertices have two branches of motion that intersect at the flat state---without losing generality, we label the unique folds 1 and 2 and the respective branches for which they are unique I and II.

{\em(ii)}  A vertex ``binds'' when one fold (or more) reaches $\pm \pi$ and prevents further motion; bound states therefore determine the ranges of the folding angles.  Figure \ref{fig:vertex_geometry}(d) shows the vertex from panel (a) in two bindings.  Fold $j$ binding creates a spherical triangle of sides $|\alpha_{j-1}-\alpha_j|$, $\alpha_{j+1}$ and $\alpha_{j+2}$ that must satisfy all three permutations of the spherical triangle inequality.  Conveniently, these can be reduced to
\begin{equation}
|\alpha_{j-1}-\alpha_{j}| \ge |\alpha_{j+1}-\alpha_{j+2}|,
\label{eq:binding_fold}
\end{equation}
where again equality corresponds to a non-generic case.
Generic vertices have two binding folds that straddle a common plate, and, once identified, the values of the remaining folding angles at binding can be calculated with spherical trigonometry \cite{inprep}.

{\em(iii)} As 4-vertices have just one continuous degree of freedom, we can pick one folding angle to parameterize the others (and later, the energy).  Huffman found implicit relationships between these angles \cite{Huffman:1976}, but we have derived the full explicit configuration equations \cite{inprep}.  We choose $\rho_1$ as our parameterizing variable, and in Fig.~\ref{fig:vertex_geometry}(b) we give example curves for the vertex shown in panel (a).  These are representative of generic vertices in that they are antisymmetric, monotonic and non-linear.

{\em Energy Landscape and Multistability:}  We
model the vertex energy with torsional springs in the folds,
\begin{equation}
E_V=\frac{1}{2}\sum\limits_{i=1} \limits^{4}\kappa_i(\rho_i-\bar{\rho}_i)^2,
\label{eq:energy}
\end{equation}
where $\{\kappa_i\}$ are spring constants ( $0 \le \kappa_i \le1$) and $\{\bar{\rho}_i\}$ are rest angles ($-\pi \le \bar{\rho}_i \le \pi$). This form is both elegant \cite{Wei:2013kn, Dias:2012kq} and experimentally valid \cite{Lechenault:2014wn, Hanna:2014}, and most importantly allows for frustration when $\{\bar{\rho}_i\}$ does not reside on any of the folding branches. 
\begin{figure}
\includegraphics[scale=1.0]{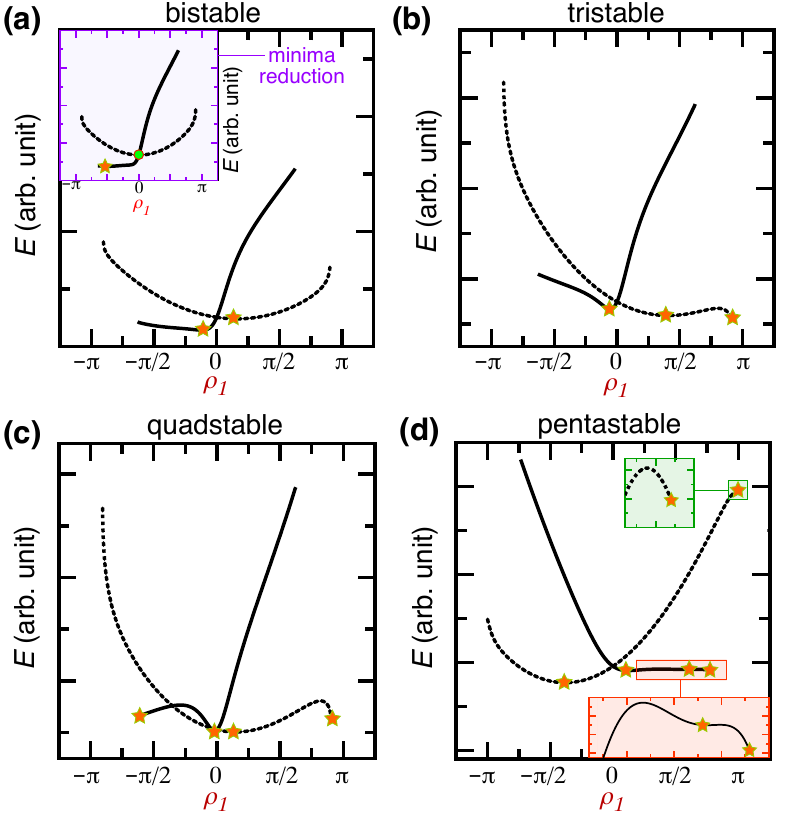}
\caption{Example bistable (a), tristable (b), quadstable (c) and pentastable (d) generic vertices.  The inset to (a) is a monostable vertex created via minima reduction, while the insets to (d) show zoomed views of the shallow minima.  For (a)-(c), $\{ \alpha_i\}$ are the same as Fig.~\ref{fig:vertex_geometry}(d), all $\kappa_i=1$ and the $\{\bar{\rho}_i\}$ are:  $\{ -1.2, 1.8, 2.3, 0.5 \}$ and $\{ -2.235..., 1.8, 2.3, 0.5\}$ for (a) and its inset, $\{ -0.1, -1.5, 2.2, 2.8\}$ for (b) and $\{ -2.6, -2.9, 3.1, 2.4 \}$ for (c).  The rare pentastable vertex has different parameters: $\{\alpha_i\}=\{1.1, 2.0, 1.9, 2\pi-5.0\}$, $\{k_i\}=\{0.3, 0.1, 1.0, 0.6\}$ and $\{\bar{\rho}_i\}=\{2.7, -2.1, -2.4, 1.4\}$.}
 \label{fig:energy_plots}
\end{figure}
The branching leads to two energy curves, thus the extreme value theorem suggests at least bistability---one minimum per branch, as in Fig.~\ref{fig:energy_plots}(a). However, as Fig.~\ref{fig:energy_plots}(b)-(d) show, it is possible to have more than one minimum per branch, leading to tri-, quad-, and even pentastable vertices.  We remark that variation in the $\{\kappa_i\}$ adds extra tunability, but is not crucial: The bi-, tri- and quadstable examples shown in Fig.~\ref{fig:energy_plots}(a-c) were created with equal strength springs and changing only the rest angles.  The rare vertex with five minima was so far only found for unequal spring constants.

To understand why multiple minima per branch occur, first note that surfaces of constant energy in the 4D space of folding angles are ellipsoidal shells centered at the global minimum $\{\rho_i\}=\{\bar{\rho}_i \}$.  Physically realizable minima occur at points where the 1D configuration curves are tangent to one of the ellipsoidal shells, and this can happen in multiple locations if the curves move both toward and then away from the global minimum \cite{Coulais:2014}. Hence, multiple minima on a single branch arise from the non-linearity in the configuration curves in conjunction with the general inaccessibility of the global minimum, {\it i.e.}~frustration.

The complexity of the configuration equations prevents us from predicting the number and location of minima analytically.  Instead, we do this numerically and discover possible stability landscapes by uniformly sampling the accessible space of $\{\alpha_i\}$, $\{\kappa_i\}$ and $\{\bar{\rho}_i\}$.  The results for generic vertices, summarized in Table I, reveal that bistable arrangements occupy the vast majority of the phase space, followed by a much smaller fraction of tristable vertices, an even smaller fraction of quadstable ones, and very rare pentastable vertices \cite{cutnote2}.  As we will show, special vertices with symmetry allow up to six stable states, and the presence of the pentastable vertex leads us to speculate that generic vertices might be capable of six minima as well---three per branch.  However, the trends in Table I make it clear that the likelihood of generating such a vertex from random sampling are incredibly small.

\begin{table}[t]
\centering
\begin{tabular}{l | c c c c c c}
{\bf geometry} & {\bf 1}  & {\bf 2} & {\bf 3} & {\bf 4} & {\bf 5} & {\bf 6}   \\ [0.5ex]
\hline
{\bf generic} &0 & 0.9311&0.0657&0.0032&$10^{-6}$&0 \\  [1ex]
{\bf flat-foldable}& 0 & 0.9418&0.0574&0.0008&0&0 \\  [1ex]
{\bf single collinear} &0 & 0.9710&0.0290&0&0&0 \\  [1ex]
{\bf double collinear}& 0 & 1&0&0&0&0 \\  [1ex]
{\bf single symmetric} &0 & 0&0&0.9768&0.0232&0 \\  [1ex]
{\bf double symmetric} &0 & 0&0&0&0&1 \\  [1ex]
\end{tabular}
\label{table1}
\caption{Multistability probabilities.  Values calculated in each case from 10$^6$ instances with uniformly sampled $\{\alpha_i\}$, $\{\kappa_i\}$, and $\{\bar{\rho}_i\}$.}
\end{table}

Monostable vertices are possible, but lie in a lower dimensional subset of measure zero and are not encountered in our sampling.  To create such a vertex, we eliminate one minimum by moving it to a branching point, {\it e.g.}~the flat state.  In the inset to Fig.~\ref{fig:energy_plots}(a), we show an example where we have tuned the parameters of a bistable vertex to have its branch II minima at the flat-state (note the similarity in the energy curves).
A vertex drawn to such a branching-point minimum will proceed to lower its energy by going to a minimum on the other branch and, as most generic vertices are bistable, such carefully tuned vertices will typically be monostable.   Conditions that guarantee this on a particular branch can be found by linearly expanding its configuration equations near the branching point, using them in the energy expression and then setting its derivative to zero (see \cite{inprep} for full details).

{\em Non-generic Vertices:  } We now turn our attention to non-generic vertices, {\it i.e.}~ones that are flat-foldable, have collinear folds, and/or have additional symmetries.  These properties influence the multiplicity and nonlinearity of the branches.

First, flat-foldable vertices meet the Kawasaki-Justin condition \cite{Demaine:2007, Kawasaki:1989, Justin:1989}, {\it i.e.}, the sum of alternating $\alpha_i$ is equal to $\pi$.  Generic flat-foldable vertices still have two branches, and their configuration curves are still non-linear, thus landscapes beyond bistability are possible.  Table I shows that flat-foldable vertices significantly suppresses the likelihood of having  more than two minima. We understand this from the observation that the configuration curves of flat-foldable vertices are, in general, less curved than those of generic vertices.

\begin{figure}[t!]
\includegraphics[scale=1.0]{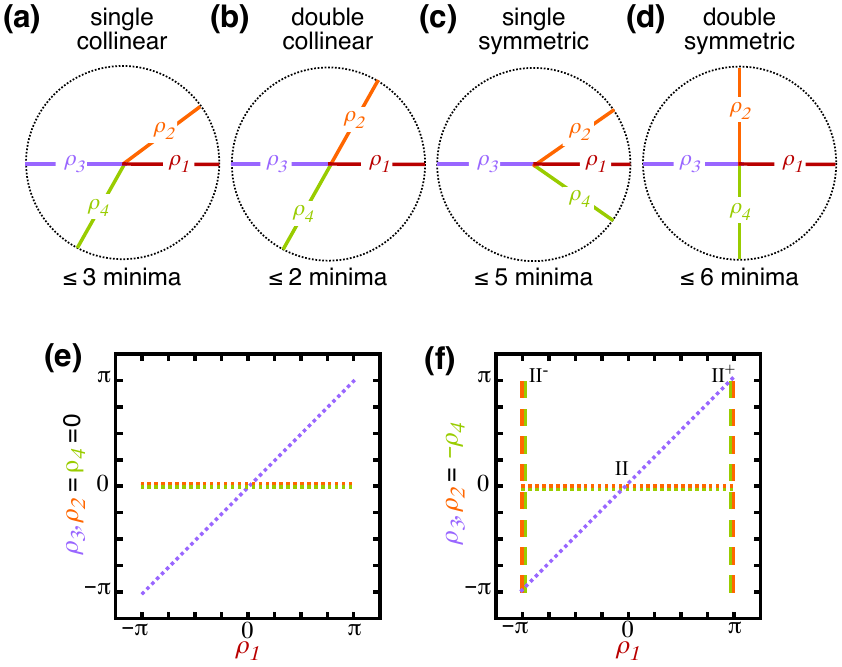}
\caption{Four types of special vertices (a)-(d).  (e)  Linear branch II curves of $\rho_{2,3,4}$ vs.~$\rho_1$ for a single collinear vertex with folds 1 and 3 collinear, as in (a).  (f)  Linear curves for branches II, II$^-$ and II$^+$ for a single symmetric vertex with folds 1 and 3 collinear, as in (c).}
 \label{fig:special_vertices}
\end{figure}

Second, single collinear vertices have two opposing folds aligned, as shown in Fig.~\ref{fig:special_vertices}(a) for folds 1 and 3. Here branch II becomes strictly linear, as in Fig.~\ref{fig:special_vertices}(e), so that the energy is simply quadratic in $\rho_1$ and yields just one minimum.  Branch I remains non-linear and can still have multiple minima.  As a consequence, single collinear vertices are more likely to have less minima than generic ones (in our sampling we find at most three---see Table I).  For a double collinear vertex [where both sets of folds are collinear but without reflection symmetry across folds --- Fig.~\ref{fig:special_vertices}(b)], both branches are strictly linear and such vertices are generically bistable unless there is minima reduction.

Third, new branches emerge when there is reflection symmetry across collinear folds [Fig.~\ref{fig:special_vertices}(c)].  For a single symmetric vertex with folds 1 and 3 collinear, we see that when they bind, subsequent folding of folds 2 and 4 becomes possible.  We designate these new branches as II$^-$ and II$^+$ (corresponding to the bindings at  $\rho_1=\rho_3=\mp\pi$, respectively).  These branches are linear, as shown in Fig.~\ref{fig:special_vertices}(f).  As with single collinear vertices, branch I remains normal and can still produce multiple minima.  In random sampling we find at most five minima (up to one for each linear branch and two for the normal one), but four minima is the most likely outcome (Table I).  For double symmetry [Fig.~\ref{fig:special_vertices}(d)], there are six linear branches and such vertices have six minima unless there is minima reduction (Table I).

{\em Metasheets:}  Before explaining how to create multistable metasheets, we first explain why such behavior has never been encountered with Miura-ora.  As the base vertex of the Miura-ori is single symmetric, it has the potential for several minima but in practice never exhibits more than one \cite{Wei:2013kn, Silverberg:2014, Hanna:2014}.  (Note we are interested in global states of rigid systems, not the local pop-through defects permitted via plate bending.)  All studies to date, however, have made the assumptions that (1) $\kappa_1=\kappa_3$ and $\kappa_2=\kappa_4$ and (2) that the global minimum lies {\it on} the normal branch.  These assumptions conspire to place the minima of each linear branch at its branching point (leading to four minima reductions), and the fact that the global minimum lies on the normal branch precludes the possibility of having a second minimum there because the configuration curves are monotonic.

\begin{figure}[t!]
\includegraphics[scale=1.0]{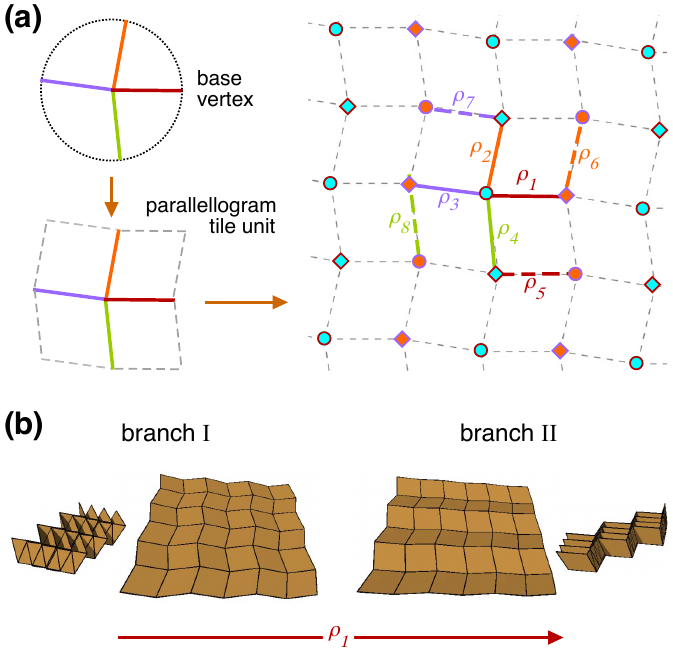}
\caption{Metasheets.  (a)   Procedure to tile an arbitrary 4-vertex.  The tiling consists of the original vertex (cyan circles), the rotated original vertex (orange circles), the complementary vertex (cyan diamonds) and the rotated complementary vertex (orange diamonds).  Homogeneous states are completely described by the folding angles $\rho_{1\rightarrow8}$.  (b)  The four homogeneous stable states of a metasheet with the same parameters as Fig.~\ref{fig:energy_plots}(c).  For movies see see \cite{supplemental_material}.}
 \label{fig:tiling}
\end{figure}

As we mentioned, the key to creating multistability lies in frustration---inaccessible minima and geometric non-linearity.  To create multistable metasheets, we take a base 4-vertex that is already multistable and build flat-state tiles by drawing parallelograms from neighboring folds, as in Fig.~\ref{fig:tiling}(a).  Although the resulting tessellation introduces three new vertices (a rotated original vertex, a ``complementary vertex'' with sector angles $\{\pi-\alpha_i\}$, and a rotated complementary vertex), the unique folds, binding folds, and binding angles of the original vertex are unchanged and the sheet remains a one-degree-of-freedom mechanism.  While the original vertex had four folding angles, homogeneous states of the tiling have eight.  Via reflection symmetry, such homogeneous states have $\rho_{i+4}=-\rho_i$.  By choosing $k_{i+4}=k_i$ and $\bar{\rho}_{i+4}=-\bar{\rho}_i$ we create a simple relationship between the sheet and base vertex energies,
\begin{equation}
E_{T}=\frac{N}{2}\sum\limits_{i=0}\limits^{4}\kappa_i\big{[}(\rho_i-\bar{\rho}_i)^2+(-\rho_i+\bar{\rho}_i)^2\big{]}=2NE_V,
\label{eq:tile_energy}
\end{equation}
where $N$ is the number of tiles in the tessellation.  

In Fig.~\ref{fig:tiling} we show images of the four minima states of a metasheet with the quadstable energy landscape of Fig.~\ref{fig:energy_plots}(c).  One particularly striking feature is the ridge patterning, which changes from vertical to horizontal as the branches are switched.  It is easy to imagine that switching between such horizontal/vertical polarizations could be useful in micromechanical devices or optical elements such as diffraction gratings.  See the supplemental material \cite{supplemental_material} for a movie illustrating the pattern transformation.

{\em Outlook:} Finally, we suggest directions for future work based on our results.  Higher $n$-vertices will lead to much richer single vertex energy landscapes---are more states possible as $n$ increases?  We have restricted ourselves to flat systems, but for non-flat systems where $\sum\alpha_i\ne2\pi$ the branching point disappears---how does the energy landscape change under these circumstances?  As pointed out by Schenk {\it et al.}~\cite{Schenk:2013kk}, folded tessellations can be stacked---is it possible to make multistable 3D folding materials?  To what degree can the energy landscapes be tuned?  With increasing complexity, one can imagine that folding planar structures might provide a platform to create metamaterials with arbitrarily tunable mechanical functionality.  

{\em Acknowledgements}  We thank I. Cohen, C. Coulais, P. Dieleman, A. Evans, R. Lang, F. Lechenault, C. Santangelo, J. Silverberg, and V. Vitelli for productive discussions.  BGC acknowledges support from FOM, and SW and MvH  support from NWO.

\bibliographystyle{apsrev}

\begin{thebibliography}{36}
\bibitem{cutnote1} This can be seen by cutting one fold --- the resulting system has three folding degrees of freedom, two of which are used to close the opened fold, leaving one non-trivial degree of freedom.
\bibitem{cutnote2} In our flat sampling of generic vertices, we found one out of $10^6$ realizations to be pentastable.  By perturbing its parameters we have found it to persist and to inhabit a finite fraction of parameter space.
\end{thebibliography}

\end{document}